\documentclass[12pt,preprint]{aastex}

\shorttitle{F Star Lithium Depletion in M35}
\shortauthors{Steinhauer \& Deliyannis}

\begin{document}

\title{WIYN/Hydra Detection of Lithium Depletion in F Stars of the Young Open Cluster M35 and
Implications for the Development of the Lithium Gap\footnote{
This is paper 21 of the WIYN Open Cluster Study (WOCS).}}

\author{Aaron Steinhauer\altaffilmark{3} and Constantine P. Deliyannis}
\affil{Astronomy Department, Indiana University, 727 E. Third St.,
	Bloomington, IN 47405-7105; aarons@astro.ufl.edu, con@astro.indiana.edu}

\altaffiltext{3}{Current address: University of Florida, Department of Astronomy,
211 Bryant Space Science Center, Gainesville, FL 32611-2055.}

\begin{abstract}

We report discovery of significant depletion of Li on the 
surfaces of F dwarf stars in the 150-Myr-old open cluster M35,
analagous to a feature in the 700-Myr-old
Hyades cluster that has been referred to as the ``Li gap.''
We have caught the gap in the act of forming: 
using high resolution, high S/N,
WIYN\footnote{The WIYN Observatory is a joint
facility of the University of Wisconsin-Madison, 
Indiana University, Yale University, and the
National Optical Astronomy Observatory.}/Hydra
observations, we detect Li in all but a few M35 F stars;
the maximum depletion lies at least $0.6-0.8dex$ below minimally 
depleted (or undepleted) stars.  The M35 Li depletion region, a) is quite wide,
with clear depletion seen from 6000K to 6700K or hotter;
b) shows a significant dispersion in Li abundance at all 
$T_{eff}$, even with stars of the same $T_{eff}$; and 
c) contains undepleted stars (as well as depleted ones) in the
(narrow) classical Hyades gap region, which itself shows no undepleted 
stars.  All of these M35 Li depletion properties
support rotationally-induced slow mixing as the
primary physical mechanism that forms the gap, and argues
against other proposed mechanisms, particularly diffusion and steady 
main sequence mass loss.
When viewed in the context of the M35 Li depletion properties, 
the Hyades Li gap may well be wider than is usually recognized.

\end{abstract}


\section{Introduction}

Abundances of lithium (Li) on the surfaces of stars have proven
to be an important diagnostic tool for astronomical investigations ranging
from cosmology to stellar structure and evolution.  
In the stellar interior, Li is burned in (p,$\alpha$) 
reactions at just a few million degrees, which makes 
its surface abundance sensitive to physical
processes occuring in the outer layers of low mass stars.
In the standard theory of stellar evolution \citep{ddk, p97}, 
the only way to affect the surface Li abundance is 
to convectively mix Li down to regions where it can burn.  
However, during the main sequence lifetime of F stars, 
the surface convection zone (SCZ) occupies only a small 
fraction (by mass) of the Li preservation region 
\citep[][hereafter DP97]{dp97}.
Thus, subsequent to the age of the Pleiades (100$\pm$20 Myr)\footnote{All
cluster ages are derived using \citet{yy01} isochrones with 0.2$H_p$ convective
core overshoot.  For discussions on the importance of overshoot see \citet{m93}, 
\citet{yy01} and \citet{RV88}.  
Overshoot ages are 25\% larger than ages without overshoot; lithium-test ages are even 
larger \citep[see][]{stauffer}.}, the standard
model predicts no Li depletion in F stars.  Indeed, F stars in the 
Pleiades and other young clusters themselves show no significant 
Li depletion, suggesting that pre-MS Li depletion in these stars, 
if any, must also have been small.
It came as a surprise, therefore, when \citet{boes86}
discovered that F stars in the Hyades (700$\pm$50 Myr) 
had dramatically depleted their surface Li abundances.  
It was immediately clear that additional
physical mechanisms beyond those usually included in the standard theory
were required to explain this phenomenon.  Subsequent studies of
open cluster Li and Be abundances are revealing a rich variety of stellar
Li depletion properties, almost {\it none} of which can be accounted
for by the standard theory \citep{d00, j00}.  
It is thus becoming increasingly important 
to ascertain what physical mechanisms are really operating inside
low mass stars, especially since there may be critical implications 
for interpreting and testing Big Bang nucleosynthesis \citep{DR04, P99}
and globular cluster ages \citep{DDP89, ddk, CDS}.

Three classes of models have emerged as possible explanations
for the F dwarf Li gap:  
1) diffusion \citep{michaud86}, where gravitational
settling and thermal diffusion drain Li out of the SCZ and surface,
2) steady main sequence mass loss \citep[][``SSD'']{ssd90}, 
where the Li preservation region is simply lost over time
(though the required mass loss rates are enormous), and
3) slow mixing driven by instabilities associated with rotation
\citep[][hereafter, C94]{pkd, pdd92, c94}
and/or gravity waves \citep{gl95, tc03}.
Rotationally-induced mixing could 
include meridional circulation and related instabilities, and/or
instabilities triggered by angular momentum loss.
To continue to try to differentiate between these scenarios 
we pursue here the novel approach of investigating the
differences in the predicted morphology of the Li gap
in its early (and heretofore uninvestigated) stages.
(See \S 3 for previously used diagnostics.)

The nearby and only moderately-reddened 
($\sim$800 pc, $E(B-V) = 0.20$), very rich open cluster M35 
has an ideal age \citep[$160 \pm 20 $Myr,][]{d04}
for studying possible early stages of Li gap formation,
soon after the age of the Pleiades.
There is also excellent membership information
from the proper motion study of \citet[][hereafter, MS86]{MS}
and WOCS radial velocity data \citep{m04}.

\section{Data and Analysis}

This section provides a brief summary of our methods.  For more
details see \citet{s04}; \citet{s03}; and 
\citet[][hereafter, DSJ02]{dsj02}.

High S/N spectra of 165 dwarf members (P generally $> 0.90$ in MS86)
of M35 were taken in the region of the 6708~\AA~Li~I resonance
doublet using three configurations of WIYN's Hydra/MOS during
three observing runs in 1997 November and December, and 1998 January.
Nearly all MS86 members in the range $V=12.25-15$ were observed.
Each star was exposed for a total of about 5 hours, yielding
S/N per 0.2~\AA pixel of 100--300 in the co-added spectra.  
(R = 18,000 = 1.9pix for the red fibers and 13,000 = 2.5pix
for the blue fibers.)

Data reduction employed standard IRAF techniques ({\it not}
dohydra).  Fits to the daytime sky spectra were used to correct
for the throughput variations between fibers and sky subtraction 
used an average of $\sim20$ sky fibers per exposure.  Effective
temperatures were determined using WOCS BVRI CCD photometry \citep{d04}.
An empirical relation was made between the $B-V$ color of the fiducial
single-star sequence and the other colors ($B-R,B-I,V-R,V-I,R-I$), and
an effective $B-V$ was calculated for each.  These
six effective $B-V$'s were then averaged, and this value was converted
into $T_{eff}$ using the $(B-V)$--$[Fe/H]$--$T_{eff}$ relation in DSJ02.
For an input [Fe/H] in that relation, we used the M35 cluster average
of $-0.143\pm0.014$ {\citep[][the error is the standard deviation of the mean, 
${\sigma}_{\mu}$; there may be additional systematic errors]{d04b,s03} } 
which agrees nicely with the value $-0.21\pm0.10$ 
from \citet[][hereafter, BDS01]{b01}.
Then, log g is taken from $Y^2$ isochrones \citep{yy01},
and $\xi$ is determined using \citet{edvard}.

The small contribution to the Li equivalent widths, $W(Li)$, from
the Fe~I line at 6707.45~\AA~ was removed using
\begin{displaymath}
W(Fe~I_{6707.45}) = (43.2 \times \langle B-V \rangle - 17.1) \times 10^{([Fe/H] -
 [Fe/H]_{Hyades})},
\end{displaymath}
which was determined empirically from high resolution and very high
S/N Hyades Li data of \citet{thdp}.  Absolute Li abundances,
$A(Li)$\footnote{$A(Li) \equiv 12 + \log (n(Li)/n(H))$.},
were determined by interpolating on the curves of
growth of S04.

The error in W(Li), ${\sigma}_W$, is derived using
the relation from \citet{dpd93}, and depends on the S/N 
and FWHM of each Li line.  Upper limits are simply 3${\sigma}_W$.
The Li line is also sensitive to $T_{eff}$, but not to log g or $\xi$:
in the gap, ${\Delta}{T_{eff}}=100K$ implies ${\Delta}A(Li)=0.08dex$.
For the WOCS photometry of our stars in a given filter, typical
${\sigma}_{\mu}$ from multiple
measures are less than 0.01mag.
Typical internal errors in the $T_{eff}$, propagated from
${\sigma}_{\mu}$ of the averaged $B-V$'s, are 10--50K for known single
stars.  The reported ${\sigma}_A$ in Table 1 has contributions
from ${\sigma}_W$ and ${\sigma}_{T_{eff}}$, added in quadrature.
Systematic errors due to $E(B-V)$ and the $T_{eff}$ scale itself
may be at least as large as 100K, but since they would affect
the whole ensemble in essentially the same (small) way, they do not
affect our conclusions. 

Following the initial selection of $P>0.9$ MS86 members,
stars were rejected if, a) they are photometric
nonmembers in {\it any} of our six CMD's, b) they are radial velocity 
non-members from our spectra or
from those spectra used in the $V_{rad}$ study, or
c) they appear to be probable binaries based on a photometric
and spectroscopic analyses.

\section{Results and Discussion}

Figure 1 shows $A(Li)$ as a function of $T_{eff}$ for the Pleiades,
M35 (this study), and the 
Hyades.\footnote{See DSJ02 for sources of Pleiades and Hyades data.}
M35 data from BDS01 are also shown.  All data from the literature
have been reanalyzed using our methods and photometry, where available, to
give the most consistent picture possible.
Several aspects of the Li--$T_{eff}$ morphology in M35
are worthy of notice.  

First, it is clear that, as compared to the nominal upper bound plateau
between 6300 and 6800K with A(Li) = 3.2, many stars between 6000 and
6700 K (or beyond) have depleted their surface Li abundances.
Although a small amount of Li depletion cannot be ruled out for the
Pleiades, the Li gap has {\it definitely} begun to form
in M35.  This is clearly illustrated in Figure 2, which shows the
continuum-normalized spectra in the Li region for four pairs of
stars.  The stars in each pair have similar $T_{eff}$, and thus it
is not surprising that their Fe I and Ca I line strengths (which
depend primarily on $T_{eff}$ and intrinsic abundance) are identical;
the Li abundances, however, are strikingly disparate.  (Note that
where necessary, spectra of slowly rotating stars were artificially 
broadened to match their pair.)  Although the errors listed in Table 1
are internal (see \S 2), we stress again that systematic errors due to 
uncertainties in $E(B-V)$ or $T_{eff}$ scale do not affect 
a differential analysis of stars at the same $T_{eff}$.
Formally, the Li differences are
significant at the $5-10\sigma$ level; the depth of the depletion far 
exceeds our errors.  The upper and lower envelope of stars
in the gap region of M35 differ by more than 0.5 dex, which corresponds to
to a factor of more than 3 in abundance (and in line strength).
Thus, M35 illustrates that the Li gap begins to form early.

Second, the Li depletion region is quite wide in Teff.  The same four
pairs illustrate that its width exceeds 700K (from 6000 to 6700K+),
which is clearly wider than the canonical Hyades Li gap (the region
6550-6700K that contains the most extreme Hyad depletions).  But
perhaps this canonical Hyades gap should be viewed as being part of
a wider depletion structure: after all, Hyad Li abundances in the
region 6300-6550K show Li dispersions of about 1 dex and are,
themselves, depleted by up to 1 dex or more.  Note also the highly
overdepleted Hyad at A(Li)~1.5,6000K in Thorburn et al. 1993.
Furthermore, it can be argued that although Praesepe has the same 
age as the Hyades, its Li gap (or, more generally, F dwarf
Li depletion region) is also quite wide, with dramatic depletion
seen all the way from 5900 to 6700K (Soderblom et al. 1993b). 
And, like the Hyades, Praesepe has a star near A(Li)~1.5, 6000K.
It may be the case that wide Li gaps are more typical than the
narrower one in the Hyades, even if that one is extended from 
6300K to 6700K as we propose here.

Third, in M35, the full $T_{eff}$ range from 6000 to 6700K shows
both undepleted (or minimally depleted) stars as well as depleted
ones.  By contrast, the Hyades Li gap has no star with $A(Li)>3.0$
in the $T_{eff}$ range 6350-6650K, and Praesepe has just a few
high-Li stars in its gap.  Also, the bottom of the M35 Li
depletion region (gap?) seems fairly uniform from 6000 to 6600K
(though note the two upper limits at 6500K).  Does the blue side
of the gap typically deplete more precipitously later on, as perhaps
suggested by the Hyades, or does significant depletion continue
throughout a wider Teff region, as perhaps suggested by Praesepe?
Is the Hyades typical or unusual?  Data from additional (rich)
clusters are needed to address this issue.

All of the results discussed above strongly support models of stellar 
evolution that include rotationally induced mixing, while arguing
against diffusion and mass loss.  In the Yale rotational models 
where mixing is in part driven by angular momentum loss,
Li dispersions at fixed $T_{eff}$ are a natural consequence of
the fact that stars form with a distribution of initial angular
momenta ($J_o$).  Stars with larger $J_o$ will eventually lose more 
angular momentum and deplete more Li.  Figure 3 shows the rotational
models of DP97 (solid lines) and the diffusion models of
\citet[][hereafter RM93, dashed lines; both interpolated slightly to an age of 150Myr]{rm93}
as compared to our M35 Li data.  The general morphology of the wide gap
is well-reproduced by the DP97 model predictions, and a reasonable distribution
in $J_o$ produces exactly the degree of dispersion that is observed.
Both minimally depleted and more significantly depleted stars are
predicted throughout the gap region.  Note the suggestion from the
models that the upper bound plateau at A(Li)=3.2 might itself be
slightly depleted.  This is consistent with the apparently slightly
higher upper bound in the Pleiades, although Galactic Li evolution
might also create slight differences in the initial cluster A(Li).

Some predictions of the rotational models of C94
are similar to those of DP97 \citep[see discussion in][]{d98}.  
\citet{tc98} improved C94's treatment of angular momentum transport
by wind-driven meridional circulation and shear turbulence and
concluded that they could explain the hot side ($T_{eff}>6700K$)
of the Li gap and other abundance anomalies in more massive stars;
however, they overdepleted Li in stars cooler than 6700K.
\citet{tc03} argue that it is plausible that
in these cooler stars angular momentum
transport is dominated instead by gravity waves (and not magnetic
fields), which have less associated mixing (and Li depletion).
Note that a distribution in $J_o$ still leads to a spread in A(Li).
We call for the construction of detailed models to
see whether this interesting approach can reproduce the features
of the young M35 Li depletion features discovered here (and of course numerous
other Li/Be/B patterns observed in low mass stars). 

Diffusion theory has difficulty explaining several aspects of our M35 
observations.  First, the depth of Li depletion is greater than the predictions of the RM93 
diffusion models which, for M35, predict a maximum depth of only 0.3 dex,
while we observe depletion of 0.6 dex or more.  The more recent
diffusion models of \citet{trm98} include complicating effects of turbulence
but this tends to dilute the effects of diffusion.
The RM93 models also predict a unique Li abundance
at a given $T_{eff}$, in obvious disagreement with the
observations presented here.  
While one might imagine parameters in the models that could
change from star to star to modulate the effects of diffusion,
no such parameter has been
suggested or explicitly modeled by the studies mentioned above.
What is most troublesome for diffusion is the overall
morphology of the Li depletion seen in M35.  At 150 Myr, the diffusion
Li gap of RM93 has its maximum depletion occur at 6900K, and is very narrow,
with a half width of about 100K.  Our data show significant depletion
over more than 700K, and extending as cool as 6000K.  (RM93 also acknowledge
that their models don't reproduce the Hyades gap very well.)  
Diffusion cannot be the dominant mechanism forming the Li gap.

Similarly, mass loss cannot be responsible for the depletion seen in Figure 1.
In order to strip away the entire Li preservation 
region of F stars of the 700 Myr 
Hyades cluster, SSD report that mass loss rates 
are required that are more than 10,000
times the current mass loss rate of the Sun.  
In order to make this proposal viable,
they conjectured that pulsationally unstable 
$\delta$ Scuti stars might be responsible
for such high rates, but these stars are only thought to extend 
as cool as about 6900K, so the coolest $\delta$ Scuti is hotter than the
hottest Li gap star.  With the discovery of significant depletion of Li in M35, 
these mass loss rates must be even higher, and 
the depletors at 6000K cannot possibly be explained by $\delta$ Scuti stars.

It is worth noting that even if not dominant, 
diffusion may yet play a role in affecting light 
element abundances of solar-type stars, since it is expected to occur in 
sufficiently stable radiative layers.  Indeed, helioseismology suggests 
the action of (downward) helium diffusion \citep{bpw95}, 
and the super-Li rich ``Li Peak'' dwarf J37 of NGC 6633 reveals the 
action of (upwards) Li diffusion \citep[DSJ02, but see][]{lg03}. 
In Fig. 1, just hotter than the A(Li)=3.2 upper bound plateau extending
from 6300 to 6800K,  note the four stars with A(Li)$\sim$3.3 between 6900
and 7200K.  It is conceivable that these somewhat high A(Li) might
somehow be related to upwards diffusion; however, the RM93-predicted Peak
is not as wide as 300K, and the Li-depleted stars in the same 
$T_{eff}$ range would require a different explanation.
Finally, although it is becoming increasingly clear that slow mixing 
affects the radiative layers below the SCZs of low mass stars,
the challenge remains to decipher what combination 
of specific physical mechanisms are at work.

We conclude that the study of the timing and morphology of the formation of
of the Li gap provides a powerful new method to study the physical mechanisms
occuring inside stars.  Results from applying this method to M35
join a variety of other evidence that points to the
action of slow mixing in low mass stars, including the Li-Be depletion 
correlation \citep{d98}, higher Li in short period tidally locked binaries 
\citep[``SPTLBs'',][]{thdp, con94, rd95},
moderately rapid Li depletion in M67 subgiants \citep{d97, sd00}, 
the existence of a moderate Be Gap \citep{bk02} though (so far) 
no B Gap \citep{b98}, the continued depletion of Li during the 
main sequence \citep{j97, d00, j02},
and the Li dispersions at fixed $T_{eff}$ (Thorburn et al.; \citealt{s93b}).

This work has been supported by the National
Science Foundation under Grants AST-9812735 and AST-0206202.

\clearpage

\begin{deluxetable}{llccc}
\tablewidth{0pt}
\tabletypesize{\scriptsize}
\tablecaption{Comparison Pair Parameters}
\tablehead{
\colhead{Star\tablenotemark{1}}            &
\colhead{$T_{eff}$}       &
\colhead{S/N/pix}         &
\colhead{$W(Li)$}         &
\colhead{$A(Li)$}}
\startdata
 149   & 6749 $\pm$  25 & 200 &   33 $\pm$    5  &  2.85  $\pm$     0.07    \\          
 212   & 6732 $\pm$  28 & 200 &   74 $\pm$    4  &  3.28  $\pm$     0.04    \\[0.15in]  
  		             		        	                            
 429   & 6559 $\pm$  19 & 170 &   24 $\pm$    4  &  2.55  $\pm$     0.09    \\          
 308   & 6525 $\pm$  22 & 150 &   84 $\pm$    5  &  3.21  $\pm$     0.04    \\[0.15in]  
  		             		        	                            
 128   & 6380 $\pm$  47 & 170 &   32 $\pm$    6  &  2.56  $\pm$     0.10    \\          
$106_r$& 6376 $\pm$   7 & 175 &   83 $\pm$    3  &  3.08  $\pm$     0.02    \\[0.15in]  
  		             		        	                            
 313   & 6010 $\pm$  60 & 150 &   43 $\pm$    6  &  2.42  $\pm$     0.09    \\          
$192_r$& 5986 $\pm$  24 &  95 &  115 $\pm$    6  &  2.98  $\pm$     0.04    \\          

\enddata
\tablenotetext{1}{star identifications from \citet{MS};  stars with $_r$ designation have been 
artificially rotationally broadenend in Figure 2 to match their pairs. }

\end{deluxetable}

\clearpage

\begin{figure}
\plotone{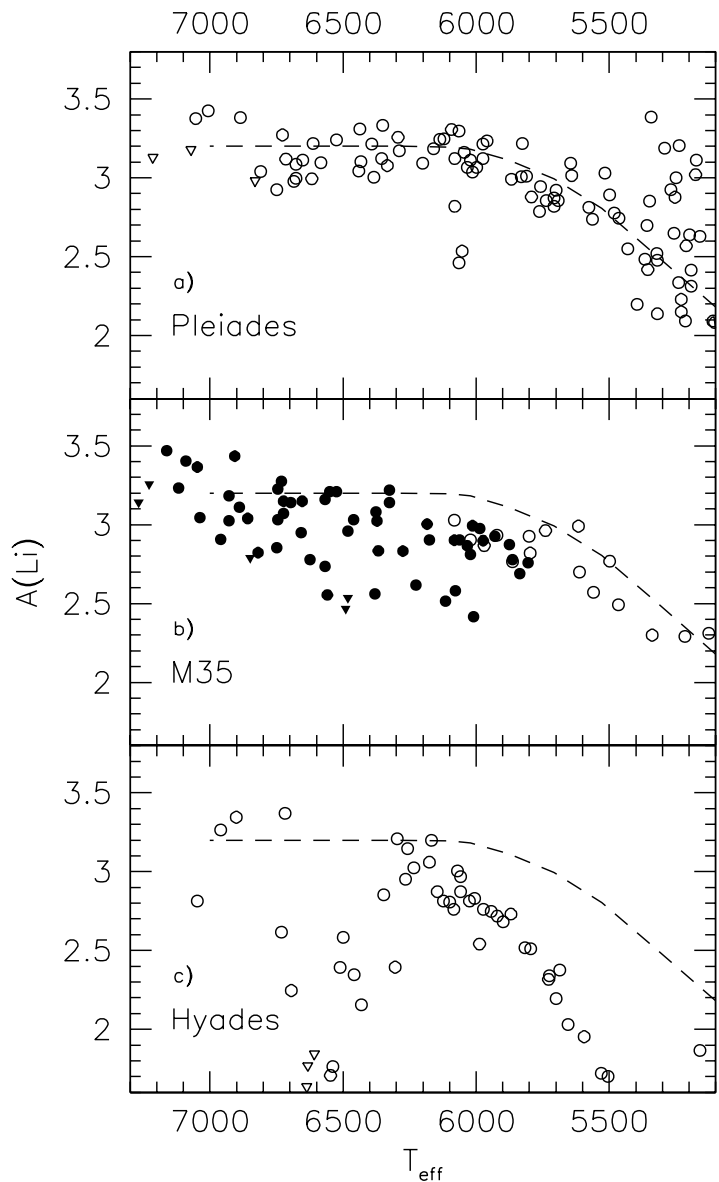}
\caption{Lithium abundances vs. $T_{eff}$ for probable single stars
in (a) the Pleiades, 
(b) M35, and (c) the Hyades.  Solid symbols are results from this
work, open symbols are data reanalyzed using methods from this paper.
Circles are detections, while triangles denote upper limits.
Dashed lines indicate the Pleiades trend.}
\end{figure}

\clearpage

\begin{figure}
\plotone{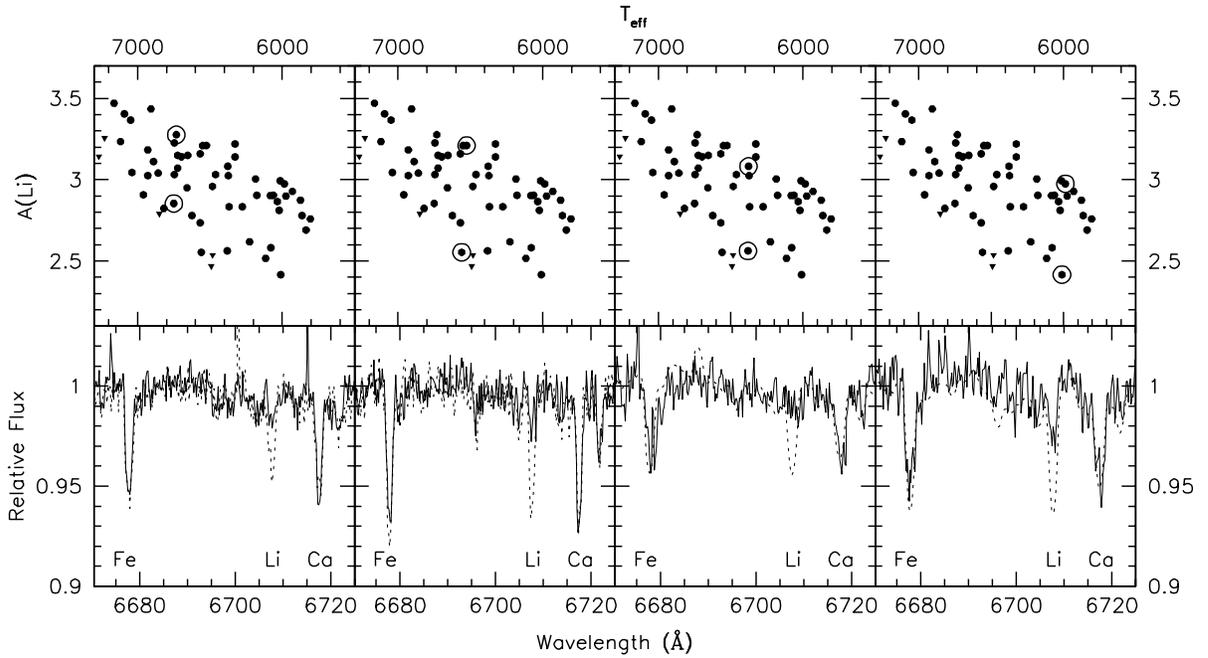}
\caption{Spectra in the Li I 6707.8~\AA~ region for four pairs of M35 dwarfs,
of similar $T_{eff}$, but different A(Li). Note that some of the spectra have
been artificially broadened to match the faster rotator.\label{fig2}}
\end{figure}

\begin{figure}
\plotone{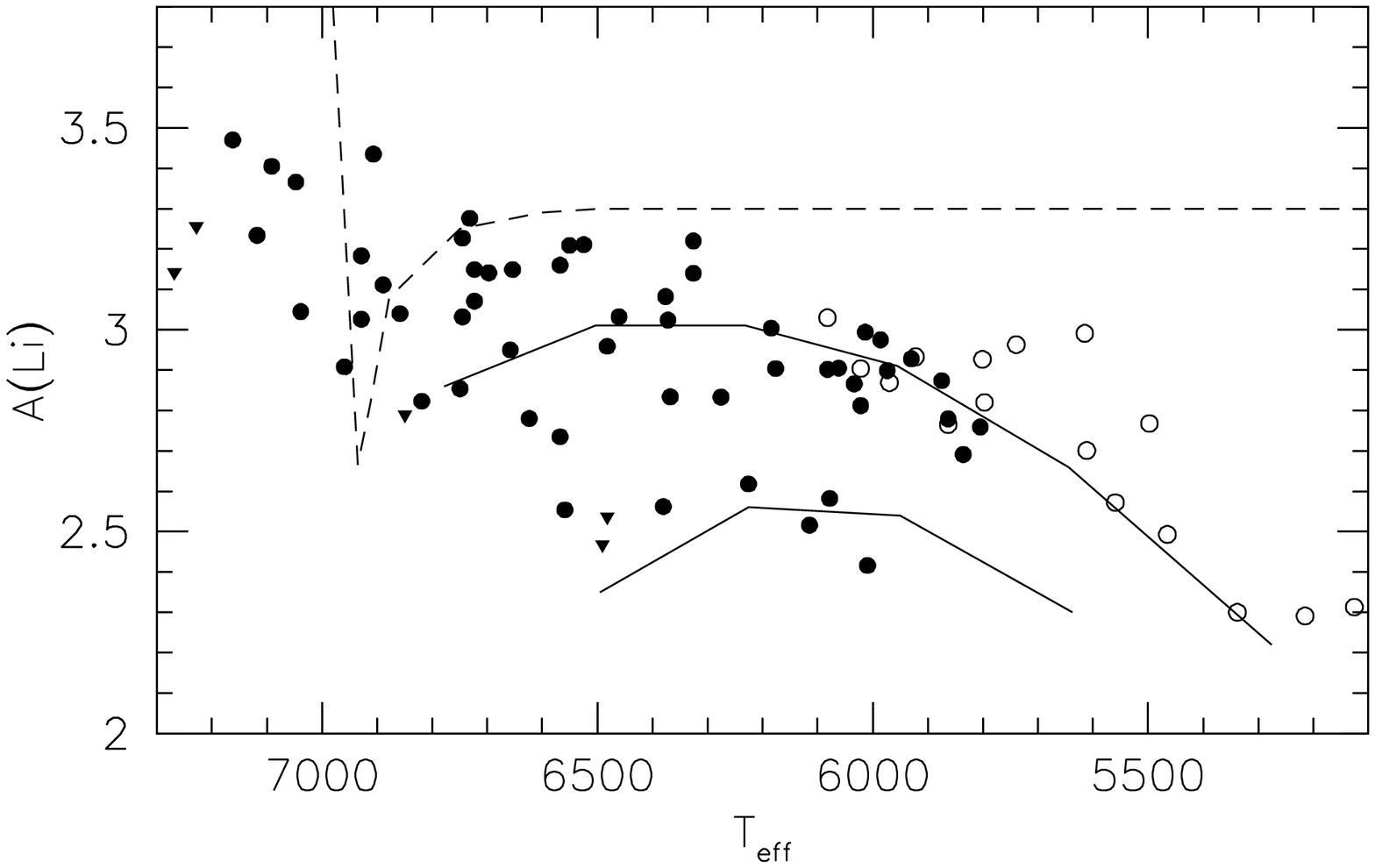}
\caption{Li abundances in M35 with solid symbols from this paper, and open
symbols from BDS01 as reanalyzed here.  The dashed line represents
predictions from the diffusion models of \citet{rm93} and the solid lines
from the rotationally induced mixing models of \citet{dp97}, with the upper
curve corresponding to $v_{init} = 10km/s$ and the lower curve to 
$v_{init} = 30km/s$.  Both models have been interpolated slightly
to 150Myr and 
assume a meteoritic value of the inital abundance ($A(Li) = 3.3$).}
\end{figure}

\end{document}